\documentclass[]{aa}
\usepackage{epsfig}
\usepackage{wasysym}
\begin{document}

\def\kms{\,km\,s$^{-1}$}       
\def\Msol{\ {\mathrm M}_\odot}             
\def\Rsol{\ {\mathrm R}_\odot}

\title{A planet-sized transiting star around OGLE-TR-122\thanks{Based on observations collected 
with the VLT/UT2 Kueyen telescope (Paranal Observatory, ESO, Chile) 
using the FLAMES+UVES spectrograph (program ID 072.C-191)}}
     
    \subtitle{Accurate mass and radius near the Hydrogen-burning limit}                        
\authorrunning{C. Melo et al.}

\titlerunning{Transiting M-dwarf around OGLE-TR-122}

 \author{   
            F. Pont\inst{1}, C. H. F. Melo\inst{2}, F. Bouchy\inst{3,4},  
	    S. Udry\inst{1}, D. Queloz\inst{1}, M. Mayor\inst{1}, N. C. Santos\inst{5}  
  }

   \institute{
          Observatoire de Gen\`eve, 51 Ch. des Maillettes, 1290 Sauverny, Switzerland
    \and
    European Southern Observatory, Casilla 19001, Santiago 19, Chile 
   \and
     Laboratoire d'Astrophysique de Marseille, Traverse du Siphon, BP 8, 13376 Marseille Cedex 12, France
     \and
         Observatoire de Haute Provence, 04870 St Michel l'Observatoire, France
   \and 
      Centro de Astronomia e Astrof\'{\i}sica da Universidade de Lisboa, Tapada da Ajuda, 1349-018 Lisboa, Portugal 
    }

  \mail{frederic.pont@obs.unige.ch}

  \date{Received / Accepted}

\abstract{

We report the discovery and characterisation of OGLE-TR-122b, the smallest main-sequence star to date with a direct radius determination. OGLE-TR-122b transits around its solar-type primary every 7.3-days. With $M=0.092\pm0.009 \Msol$ and
$R=0.120^{+0.024}_{-0.013} \Rsol$, it is by far the smallest known eclipsing M-dwarf.  The derived mass  and
radius for OGLE-TR-122b are in agreement with the theoretical expectations. OGLE-TR-122b is the first observational evidence that stars can indeed have radii
comparable or even smaller than giant planets. In such cases, the photometric signal is exactly that of a transiting planet and the true nature of the companion can only be determined with high-resolution spectroscopy. 
}
  \authorrunning{Pont et al.}
\maketitle
\section{Introduction}

More than two dozen photometric searches for planetary transits are currently under way. However, up to now the OGLE planetary transit survey (Udalski et al. 2002a) has been the only one able to detect transit signals of the order of 1 \%, typical of Jupiter-sized planets with solar-type hosts. The OGLE planetary transit survey has monitored several fields in the Galactic disc (Udalski et al. 2002a,b,c, 2003, 2004), identifying 177 planetary transit candidates. Subsequent radial velocity follow-up of these candidates has resulted in the discovery of five new transiting exoplanets (Konacki et al. 2003, Bouchy et al. 2004, Pont et al. 2004, Bouchy et al. 2005, Konacki et al. 2005). 

Most OGLE transit candidates actually turned out to be eclipsing binaries, most commonly small  M-dwarfs transiting in front of F-G dwarfs. These objects provide a very interesting by-product of the planetary transit search: precise masses and radii for small stars down to the brown dwarf domain. So far, there is scant observational constraints on the mass-radius relation of very low-mass stars. Very light stars are known to exist (e.g. Close et al. 2005), but below $0.3 \Msol$, radii are known only for one eclipsing binary (CM Dra, Metcalfe et al.~\cite{metaetal96}) and three nearby M-dwarfs measured in interferometry (GJ191, GJ551, S\'egransan et al.~2003, and GJ699, Lane et al.~2001).

Modeling of the interior and the evolution of very low-mass stars (i.e., stars close
to the H-burning limit and below) is a complex issue because it strongly depends on the
equation of state, whose derivation for such low temperatures as found inside
these objects requires a detailed description of strongly
correlated and partially degenerate classical and quantum plasmas (see review by Chabrier \& Baraffe~\cite{chabar00}).
In this context, accurate measurements of mass and radius are valuable observational constraints
since both reflect the physical properties characteristic of the interior of these objects.

In 2003 and 2004, we have followed in high-resolution spectroscopy, with FLAMES/UVES on the VLT, 60 of the OGLE transiting candidates (Bouchy et al. 2005, Pont et al. 2005). Additionally to the characterisation of 5 transiting exoplanets, this work has lead to the discovery of seven transiting M-dwarfs with $M\leq 0.3 \Msol$. 

In this letter, we present the new eclipsing binary OGLE-TR-122, a system composed of a solar-type primary and a very low mass secondary close to the Hydrogen-burning limit -- by far the smallest main-sequence star with a precise radius measurement to date. We describe the determination of the parameters of the system (Sect.~\ref{analysis}) and briefly comment on the position of the object in the mass-radius diagram (Sect.~\ref{mrr}). Finally, we point out in Sect.~\ref{det} the implications of this discovery for photometric transiting planet searches.

\section{Analysis of the spectroscopic  and lightcurve data} 
\label{analysis}
The spectroscopic observations were collected in March 2004 
with the multi-object spectrograph facility FLAMES+UVES (Pasquini et al.~\cite{pasetal02}) attached to
the VLT UT2-Kueyen telescope, at Paranal Observatory, ESO. 
OGLE-TR-122 was observed 6 times as part of our spectroscopic follow-up of
OGLE planetary transiting candidates in Carina (see  Pont et al. 2005).

\begin{table}[ht!]
\caption{Radial velocity measurements and cross-correlation parameters for OGLE-TR-122. Date are from BJD=2453000.}
\label{table:rv}
\begin{tabular}{lrrrrr}
\hline\hline
 Date & RV & depth & FWHM & $S/N$ & $\sigma_{RV}$ \\
$\,$[BJD] & [\kms] & [\%] & [\kms] & &  [\kms]\\ 
\hline
78.634770  & $-$7.720  & 17.54  & 11.2  & 3.7  & 0.159 \\
79.573374  & $-$2.620  & 21.85  & 11.4  & 5.3  & 0.094 \\
80.589249  & 2.930   & 19.22  & 11.8  & 4.7  & 0.119 \\
81.690009  & 7.820   & 21.48  & 11.9  & 5.3  & 0.097 \\
84.579819  & $-$9.535  &  26.78 & 12.2  & 7.5  & 0.063 \\
85.678956  & $-$8.809  &  25.92 & 12.1  & 7.3  & 0.065 \\


\hline
\end{tabular}
\end{table}

\begin{figure}[ht!]
\resizebox{\hsize}{!}{\includegraphics{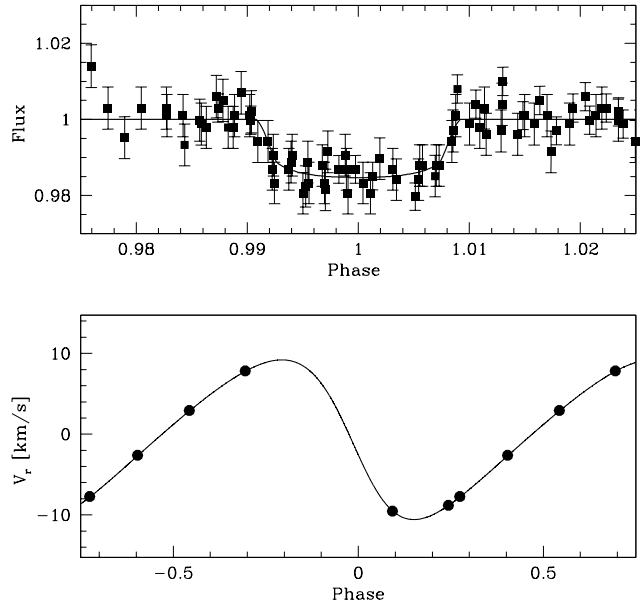}}
\caption{ 
{\it Top.} phase-folded light curve and best-fit transit curve for OGLE-TR-122.
{\it Bottom.} Radial velocity measurements and orbital solution.}
\label{fig:orbit}
\end{figure}

The radial velocities for OGLE-TR-122 are given in
Table~\ref{table:rv}. The radial velocity data and its best-fit orbital solution are shown in Fig.~\ref{fig:orbit}, together with the OGLE light curve and our transit solution. In the orbital solution, the period and epoch of the transits were fixed to the value given by the lightcurve. The cross-correlation signal is deep and moves by more than its width in our measurements with no significant change of shape, indicating that there is no significant second component blended in the spectra. 

\begin{figure}[t!]
\resizebox{\hsize}{!}{\includegraphics{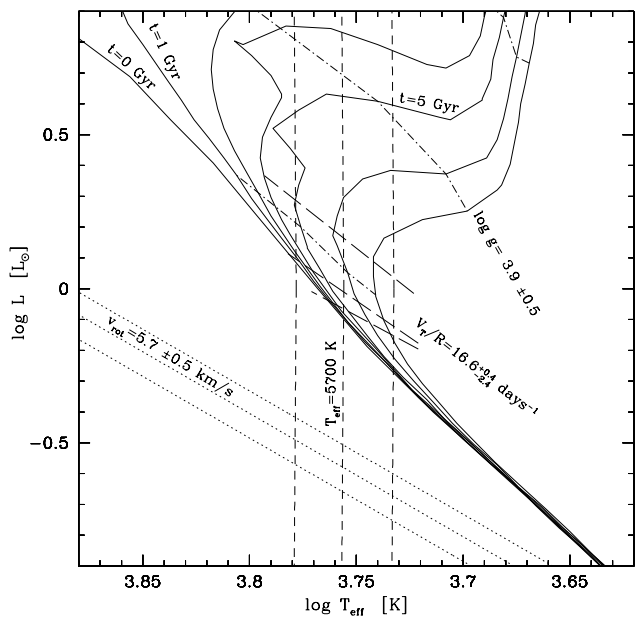}}
\caption{Illustration of the different constraints on the primary of OGLE-TR-122 compared to stellar evolution models from Girardi et al. (2002) with $Z=0.02$  for ages 0, 1, 2, 3, 5, 10 and 15 Gyr (thin lines). Long-dashed lines show the contraints from $V_T/R$ (related to the transit duration), short-dashed lines from $T_{\rm eff}$, dash-and-dot lines from $\log g$. Dotted lines shows the isoradius lines corresponding to pseudo-synchronised rotation. In each case the lines show the central value and $\pm$1-sigma interval.}
\label{fig:cmd}
\end{figure}

Observational constraints from the radial velocity orbit, transit shape and spectral analysis were combined  as described in Bouchy et al. (\cite{bouchyetal05}) to yield the physical characteristics of the two bodies involved in the transit. The relative mass and size of the secondary are  obtained from the transit depth and radial velocity orbit (Fig.~\ref{fig:orbit}), with $R_c/R=0.114 \pm 0.008$ and $f(m)=6.73\pm 0.19 \cdot 10^{-4}$, where $R_c$ and $M_c$ are the radius and mass of OGLE-TR-122b and $f$ the mass function. The primary mass and radius are determined from the spectroscopic parameters ($T_{\rm eff}, \log g$, $[Fe/H]$) and the duration of the transit (related to $M$, $R$, $i$ and $P$ through the equations of Keplerian motion).  Because the orbit is eccentric, the equations relating the physical parameters to the transit duration and the radial velocity semi-amplitude have to be modified accordingly. These constraints are applied to the locus defined for stellar parameters by the stellar evolution models of Girardi et al. (2002) to determine the likely values of $M$ and $R$, as illustrated in Fig.~\ref{fig:cmd}.

Close binaries undergo tidal forces that tend to synchronize the rotation of the components and to circularize the orbit. For late-type primaries the transition is observed in the 5-10 day range. Using the theoretical relations of Zahn (1977) for stars with convective envelopes, the circularisation timescale for OGLE-TR-122 is 50 Gyr, and the synchronisation timescale is 1 Gyr. The measured orbit of OGLE-TR-122 indicates that it is not yet circularized, as expected from the large timescale. The synchronisation timescale is comparable to the age of disc stars, so that rotation of the primary may or may not be synchronised. For elliptic orbit, the system is expected to move towards "pseudo-synchronisation", with the rotation velocity tuned to the velocity of passage at periastron (Zahn 1977). Given the measured rotation velocity of OGLE-TR-122 ($v \sin i = 5.7 \pm 0.5$ \kms), assuming pseudo-synchonisation would imply a primary radius of $R=0.67\pm 0.06 \Rsol$, in conflit with constraints from the spectroscopic parameters and transit duration (see  Fig.~\ref{fig:cmd}). We conclude that OGLE-TR-122 is not pseudo-synchronised yet. Note that it is likely that its rotation velocity has increased already under the influence of the close companion and is approaching synchronisation. This makes lower radii more likely than higher radii for the primary and the companion, because they imply a measured rotation velocity nearer to synchronisation. 

In Table 2 we summarize the parameters obtained for the OGLE-TR-122 system and its components. A coherent solution is found for $R$ and $M$, but additional measurements of the spectroscopic parameters and the transit shape would certainly be useful to decrease the uncertainty on the solution, and consequently the uncertainties on the mass and radius of the companion. The rotation velocity in Table 2 is measured from the width of the spectral cross-correlation function.




\begin{table}[hb!]
\caption{Parameters for OGLE-TR-122 and its transiting companion. $T_{\rm tr}$ and $T_{\rm per}$ are the epochs of the transits and periastron respectively.  }
\label{table:results}
\begin{tabular}{lr}
\hline
\hline
Period (days)			&	7.26867 [fixed]	\\
Radius ratio                    &       0.114 $\pm$ 0.008 \\
$V_T/R$	(days$^{-1}$ $\Rsol^{-1}$)&	16.59$^{+0.4}_{-2.4}$\\
$i$	(degrees)	                &       88-90 \\
$T_{\rm tr}$ 	(BJD)		&	2452342.28 [fixed]\\
$T_{\rm per}$ 	(BJD)		&	2452342.41$\pm$0.02\\
$e$				&	0.205$\pm$0.008	\\
$\gamma$ (km s$^{-1}$)		&	$-$0.494$\pm$0.068	\\
$K$ (km s$^{-1}$)		&	9.642$\pm$0.088	\\
$\omega$ (degrees)		&	99.2$\pm$0.8	\\
$O-C$ (km s$^{-1}$)		&	0.057	\\
$v\, {\sin i}$ (\kms)              &       5.7 $\pm$ 0.6  \\
\\
$T_{\rm eff}$ (K)			&	5700$\pm$300	\\
$\log g$			&	3.9$\pm$0.5	\\
$[{\rm Fe/H}]$			&	0.15$\pm$0.36	\\
$R\ (\Rsol)$	&	$1.05^{+0.20}_{-0.09}$	\\	
$M\	(\Msol)$	&	0.98$\pm$0.14	\\
\\
$R_c\ (\Rsol)$	& 	$0.120^{+0.024}_{-0.013}$	\\
$M_c\	(\Msol)$	&	0.092$\pm$0.009	\\
\hline
\end{tabular}
\end{table}

\section{The mass-radius relation for low-mass stars, brown dwarfs and planets}

\label{mrr}

\begin{figure}[t!]
\resizebox{\hsize}{!}{\includegraphics{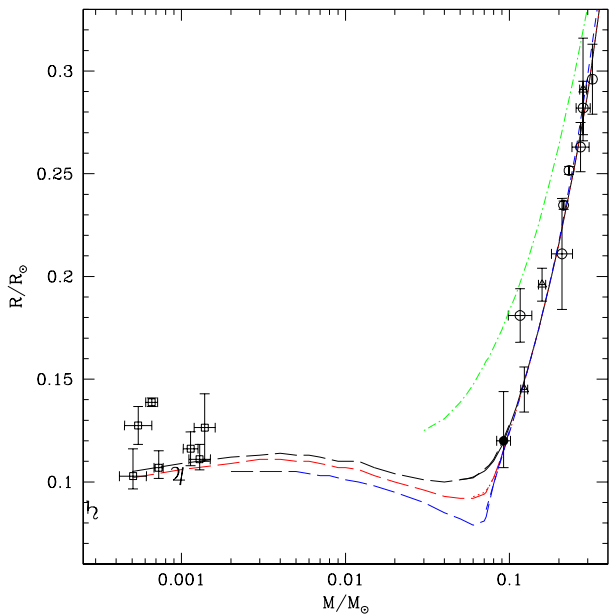}}
\caption{Mass vs. radius for observed low-mass stars and giant planets and theoretical isochrones.
 Eclipsing binaries are shown as circles (OGLE-TR-122b in black), 
 interferometric data  as open triangles. 
 Baraffe et al. (\cite{bcah98}) isochrones for masses from $0.06M_\odot$ to $1.4M_\odot$
 are plotted for 5 Gyr (solid) and 0.1 Gyr (dash-dotted). 
 Dashed lines represent the Baraffe et al. (\cite{bcah03}) COND models for
 masses for 0.5, 1 and 5 Gyr, from top to bottom.}
\label{fig:mr}
\end{figure}

Fig.~\ref{fig:mr} shows the position of OGLE-TR-122b in the mass-radius diagram compared to the other low-mass M dwarfs and giant planets with known mass and radius. CM Dra  (Metcalfe et al.~\cite{metaetal96}),  OGLE-TR-5, OGLE-TR-7 (Bouchy et al. \cite{bouchyetal05}) and OGLE-TR-72, OGLE-TR-78, OGLE-TR-106, OGLE-TR-125 (Pont et al. \cite{pontetal05}) are eclipsing binaries with radii and masses determined by combined  radial velocity  and
light curve analysis. The remaining three stellar objects (GJ191, GJ551, GJ699) were measured in interferometry and the angular radius was converted to linear radius using Hipparcos parallaxes, and masses were estimated from the $K-$band mass-luminosity relation (S\'egransan et al.~\cite{segetal03}). 
Also plotted are theoretical isochrones from Baraffe et al.(\cite{bcah98})  and Baraffe et al. (\cite{bcah03}).  

The models predict that main-sequence objects reach, near $0.07-0.1 \Msol$, Jupiter-size radii, then 
the internal pressure is described by a mixture between
the classical perfect gas+ion case and
the fully degenerate electron gas, and consequently $R \propto R_0m^{-1/8}$, i.e., the radius remains almost constant
 (Chabrier \& Baraffe~\cite{chabar00}). OGLE-TR-122b is the first object to prove empirically that main-sequence stars do indeed reach the size of gas giant planets. The position of OGLE-TR-122b is in agreement with the model predictions for $\tau\geq0.5$ Gyr.

\section{Implications for planetary transit surveys}

\label{det}


Existing surveys have found that eclisping binaries are the major source of contamination in the search for planetary transits. In many cases, features of the lightcurve, combined with some low-resolution information on the primary (broadband photometry or low-resolution spectroscopy), can be used to distinguish eclipsing binaries  from transiting planets. If the object is a grazing eclipsing binary with two components of comparable size, the shape and duration of the transits are markedly different from a planetary transits. If the companion is an M dwarf, it is usually larger than a planet, and therefore produces a deeper eclipse for a given primary radius. For short-period M dwarfs, tidal deformation of the primary produces detectable modulations of the lightcurve outside the transit. 

In this context, OGLE-TR-122 empirically  establishes two important points:
first, that some stellar objects can produce precisely the same photometric signal as transiting gas giants, and second, that these objects are not rare. OGLE-TR-122b has exactly the same radius as a planet, and therefore the shape and duration of the transit signal that it produces is strictly equivalent.  By extrapolating from the observed amplitude for OGLE-TR-5 ($P=0.8$ days, $M_c\simeq 0.27$ M$_\odot $, modulation of 7.2 mmag, see Bouchy et al. 2005) with the theoretical dependence $\sim M_c/P^2$, we estimate that the photometric variations caused by the tidal deformation of the primary are of the order of 0.06 mmag for OGLE-TR-122. Such a signal is far below the detection threshold of transit surveys. The depth of the secondary eclipse (when the M dwarf is hidden behind the primary star) is 0.02 mmag in $V$ and 0.30 mmag in $I$ for a mass of $M_c=0.09 \Msol$ and an age of 5 Gyr (using the Baraffe et al. \cite{bcah98} models for the mass-luminosity relations). This is also below the threshold of ground-based observations. In the infrared, it could be  within reach of space-based observations if the target is bright enough. However, with a slightly lower mass the object is predicted to become much fainter. For instance with $M_c=0.08 \Msol$, the amplitude of the secondary eclipse signal in $I$ becomes 0.04 mmag for an age of 5 Gyr. Therefore, in practice, objects like OGLE-TR-122 can be totally undistinguishable from a planetary transit in photometry alone.

The detection of OGLE-TR-122b provides an indication of the abundance of such objects in transit surveys, compared to gas giant planets. Three transiting planets where found in the OGLE Carina fields (see Introduction for references). Two of these have very short periods, and their detections was highly favoured by the time sampling of the survey (see e.g. Pont et al. 2005). For statistical interpretation, OGLE-TR-122b should thus be compared only with the single planet detected at $P>2$ days,  OGLE-TR-111b. Therefore the abundance of $R_c\sim 0.1 \Rsol$ companions is of the same order of magnitude as the abundance of hot Jupiters. Indeed, this statement is conservative, because in our survey we rejected many eclipsing binaries without determining their masses precisely, and there could still be other very small stellar objects in the sample. 



The implication for transit surveys is that it is not possible to be confident in the planetary nature of a transiting Jupiter-sized object from the photometric signal alone. It is not even possible in a statistical sense, because $R\sim 1$R$_{\rm J}$ transiting {\em stellar} companions are about as numerous as transiting hot Jupiters. Therefore, follow-up in high-resolution spectroscopy is essential, and candidates that are too faint for high-resolution spectroscopic follow-up are destined to remain in an undetermined "limbo" as to the nature of the transiting body. 


\begin{acknowledgements} 

We are indebted to I. Ribas for useful comments that helped improve the analysis. We would like to thank the Paranal Observatory staff for support at UT2/Kueyen.


\end{acknowledgements}

\end{document}